\documentclass[12pt]{article} 
\usepackage{graphicx,subfigure,latexsym}

\newcommand{\be}{\begin{equation}}
\newcommand{\ee}{\end{equation}}
\newcommand{\bear}{\begin{eqnarray}}
\newcommand{\eear}{\end{eqnarray}}
\newcommand{\ba}{\begin{array}}
\newcommand{\ea}{\end{array}}

\begin{document}

\begin{titlepage}
\vfill
\begin{flushright}
{\normalsize RBRC-1011}\\
\end{flushright}

\vfill
\begin{center}
{\Large\bf Flows and polarization of early photons with magnetic field at strong coupling}

\vskip 0.3in

Ho-Ung Yee \footnote{e-mail:
{\tt hyee@uic.edu}}
\vskip 0.15in

{\it Department of Physics, University of Illinois, Chicago, Illinois 60607}\\[0.10in] {\it and}\\[0.10in]
{\it RIKEN-BNL Research Center, Brookhaven National Laboratory,}\\
{\it Upton, New York
11973-5000}\\[0.15in]
{\normalsize  2013}

\end{center}

\vfill

\begin{abstract}

Recent experimental results from {\bf RHIC} and {\bf LHC} on hard photon emission rates in heavy-ion collisions indicate a large azimuthal asymmetry of photon emission
rate parameterized by the elliptic flow $v_2$. Motivated by a recent proposal that the early magnetic field created by two colliding heavy-ions may be responsible for this large azimuthal asymmetry of photon emission rate, we compute the azimuthal dependence of the photon emission rate from a strongly coupled finite temperature plasma with magnetic field in the framework of gauge/gravity correspondence. We also propose and compute a new observable, "in/out-plane polarization asymmetry", constructed from the polarization dependence of the photon emission rates. We observe that both the azimuthal and polarization asymmetry of photon emissions are strongly affected by the triangle anomaly (chiral anomaly) for low frequency regime below 1 GeV.

\end{abstract}

\vfill

\end{titlepage}
\setcounter{footnote}{0}

\baselineskip 18pt \pagebreak
\renewcommand{\thepage}{\arabic{page}}
\pagebreak

\section{Introduction}

The quark-gluon plasma created in relativistic heavy-ion collisions is an interesting new state of matter where one can test
various theoretical ideas of QCD. The challenge is that this plasma lives for very short period of 10 fm, and there are many indications that it is strongly coupled, making perturbative computations very difficult.

Their rapid evolution in time after about 1 fm is believed to be described fairly well by viscous hydrodynamics, but
the earlier dynamics still remains as an important open problem. Once hydrodynamics sets in, the data needed to describe the system reduces dramatically to a few transport coefficients, and this is both a power and limitation of hydrodynamics.
As the system is very likely strongly coupled, there has been much research on using gauge/gravity duality or AdS/CFT correspondence to describe the hydrodynamic regime of strongly coupled plasma, essentially computing the transport coefficients
non-perturbatively \cite{Kovtun:2004de}. Using gauge/gravity duality for early-time dynamics has also been pursued while it requires some amount of numerical analysis to solve relevant non-linear Einstein equations \cite{Chesler:2010bi,Heller:2011ju,Wu:2011yd,Lin:2006rf,Balasubramanian:2010ce}.

Some experimental probes see the created quark-gluon plasma in more refined way than the hydrodynamics description.
A good example is the photon emission spectra from the plasma, especially at high frequency (and momentum) whose scale is above the temperature scale of the plasma. In linear response framework which is suitable for electromagnetic probes such as photon emission, the emission spectra mirror the Green's functions of vector current that the electromagnetism couples to.
These Green's functions at high frequency-momentum are properties of the plasma beyond hydrodynamics, and they do depend on
the microscopic details of the theory. 

This break-down of universality that has been acclaimed for hydrodynamics is an important aspect one has to bear in mind in applying AdS/CFT correspondence to photon emissions, because for example N=4 Super Yang-Mills is very different from QCD microscopically. 
There is however another sense of universality that may originate just from the strongly coupled nature, and this still
remains as a useful argument for applying AdS/CFT correspondence even for probes beyond hydrodynamics such as photon emission.

Once photons are emitted from the plasma, they rarely interact with the plasma again because of the weakness of electromagnetic coupling. This aspect makes the photons ideal probes to the early dynamics of heavy-ion collisions, and there have been important
experimental results from {\bf PHENIX} \cite{Adare:2011zr} and {\bf ALICE} \cite{Lohner:2012ct} on photon emissions. At present, the amount of photons observed experimentally tends to exceed the theoretical predictions, but theoretical computations are being refined to be closer to the experiments \cite{vanHees:2011vb,Dion:2011pp}.
What is challenging is that the observed photons (both direct and indirect) are from the entire history of the heavy-ion collisions, although different regions of frequency should be sensitive to different parts of the history. The low frequency regime will naturally be sensitive to somewhat later stage of plasma evolution with lower temperatures as well as hadronic regime after-burner phase. The higher frequency spectrum is expected to get dominant contributions from early stages, but
the dynamics before 1 fm is still not very well understood.

One striking recent result from {\bf PHENIX} \cite{Adare:2011zr} and {\bf ALICE} \cite{Lohner:2012ct} is the large elliptic flow of direct photons at relatively large transverse momentum
$p_T>1$ GeV. At $p_T=2$ GeV, it is as large as $v_2=0.25$. Contrary to hadronic elliptic flow which builds up its magnitude
through collective hydrodynamic evolution, high frequency photons coming from early stage are not expected to 
be sensitive to this late time collective evolution. (The low frequency spectra do get some effects from the Doppler shifts.)
Given this, it is natural to seek a more direct source of azimuthal asymmetry relevant for the photon emission at the early
stage of the heavy-ion collision. 

An interesting candidate is the magnetic field created by two fast moving colliding heavy-charged nuclei \cite{Kharzeev:2007jp}.
Its magnitude was estimated to be as large as $eB=10^{17}$ Gauss $\sim m_\pi^2$. 
Possible experimental signatures of the effects of this magnetic field have been suggested previously in conjuction with triangle anomaly; chiral magnetic/separation effects \cite{Kharzeev:2007jp,Fukushima:2008xe,Son:2004tq} and chiral magnetic waves \cite{Kharzeev:2010gd,Newman:2005hd}, leading to charge dependent two-particle correlations \cite{Voloshin:2004vk} and charge-dependent elliptic flows of pions \cite{Burnier:2011bf,Burnier:2012ae}. See also Ref.\cite{Gorbar:2011ya}. These predictions are indeed observed experimentally at {\bf STAR/PHENIX} \cite{Abelev:2009ac,Adamczyk:2013kcb,Wang:2012qs,Ke:2012qb} and {\bf LHC} \cite{Selyuzhenkov:2011xq}, and more experimental results are on the way.

The idea that the early magnetic field may affect photon emissions leading to azimuthal asymmetry has been previously
discussed in Refs.\cite{Tuchin:2010gx,Tuchin:2012mf,Basar:2012bp,Fukushima:2012fg,Bzdak:2012fr}, mainly in the weak coupling regime \cite{Tuchin:2010gx,Tuchin:2012mf} or including conformal anomaly \cite{Basar:2012bp}. In our present study, we look at the effect of magnetic field on the photon creation, especially
its azimuthal asymmetry that leads to observable mock-up of elliptic flow, in a strongly coupled finite temperature plasma in the framework of gauge/gravity correspondence.
See Refs.\cite{CaronHuot:2006te,Mateos:2007yp} for earlier holographic computations for the photon emission rate without magnetic field, and Refs.\cite{Rebhan:2011ke,Patino:2012py} for the photon emission with an anisotropy introduced by an external axionic perturbation \cite{Mateos:2011ix}. Ref.\cite{Mamo:2012zq,Bu:2013vk} computed photon emission rates with magnetic field, but the detailed elliptic flow has not been computed. 
As the photon emissions are sensitive to microscopic details of the theory, we choose to work with the model by Sakai and Sugimoto \cite{Sakai:2004cn} which is dual to the field theory which should be closest to the real QCD in quenched approximation.

Our results indicate that the azimuthal dependence of the photon emission rates is more complicated than the simple elliptic flow pattern, which necessitates looking at higher moments of the azimuthal dependence, such as the quadrupole flow $v_4$.
We predict an interesting pattern of their ratios, $v_4\over (v_2)^2$, as a function of frequency, which might be relevant
experimentally.

We also compute polarization asymmetry of the photons emitted in the presence of the magnetic field.
After defining an observable, ``in/out plane polarization asymmetry'', $A^{\rm I/O}$, with respect to the reaction plane, we compute it as a function of frequency. Although it seems challenging to measure polarizations of photons in current experiments, it might be an interesting direction to refine the experimental photon measurements.

In both computations, we observe that triangle anomaly of the chiral symmetry plays an indispensible role;
in gauge/gravity duality it manefests as a 5-dimensional Chern-Simons term which does affect the equations of motion
in the presence of magnetic field.
Essentially the effects can be summarized as coming from the new transport phenomenon of chiral magnetic waves of chiral charges.
We explicitly check that results with and without Chern-Simons term differ
substantially to each other, which indicates that any computation without properly including chiral anomaly might be questioned.

 In our analysis, we assume a static plasma and a constant magnetic field.
In comparison to experiments, one has to integrate our results over the realistic time history of the heavy-ion collisions.
This procedure is based on an additional assumption of adiabaticity.
We leave further refinement of our study including time-evolution of the plasma to the future.

\section{Photon emission rates from equilibrium plasma}

In this section, we summarize photon emission rate formula and useful relations between Green's functions for the case of
translationally invariant thermal plasma. Note that we will not assume rotational isotropy that is broken by the presence of a magnetic field.
The differential photon emission rate summed over all polarization states from a plasma in thermal equilibrium is given by
\be
d\Gamma_\gamma = {d^3 k\over(2\pi)^3}{e^2\over 2 |\vec k|}\eta^{\mu\nu}G_{\mu\nu}^<(k)\Bigg|_{k^0=|\vec k|}\quad,
\ee
where $G^<$ is the Wightman function defined by
\be
G^<_{\mu\nu}(k)\equiv\int d^4 x\, e^{-i k x}\langle J_\mu^{EM}(0)J_\nu^{EM}(x)\rangle\quad.
\ee
The emission rate for a particular polarization state specified by $\epsilon^\mu$ is
\be
d\Gamma_\gamma\left(\epsilon^\mu\right) = {d^3 k\over(2\pi)^3}{e^2\over 2 |\vec k|}\epsilon^\mu \epsilon^{\nu *}G_{\mu\nu}^<(k)\Bigg|_{k^0=|\vec k|}\quad.
\ee
Our metric convention is $\eta=(-,+,+,+)$.
We would like to relate $G^<$ with the retarded Green's function $G^R$ defined by
\be
G^R_{\mu\nu}(k)\equiv(-i) \int d^4 x\, e^{-i k x}\,\theta(x^0)\,\left\langle \left[J_\mu^{EM}(x),J^{EM}_\nu(0)\right]\right\rangle\quad,\label{GR}
\ee
since $G^R$ is naturally computed in holography with incoming boundary condition on the black-brane horizon as we will see in the next section.
It is convenient to define
\be
G^>_{\mu\nu}(k)\equiv -\int d^4 x\,e^{-ikx}\langle J_\nu^{EM}(x) J_\mu^{EM}(0)\rangle\quad,
\ee
and using Lehmann representation one can easily derive the relation
\be
G^>_{\mu\nu}(k)=-e^{\beta k^0} G^<_{\mu\nu}(k)\quad.
\ee
On the other hand, from definitions of $G^{>,<}$ one has
\be
G^N_{\mu\nu}(k)\equiv\int d^4 x\,e^{-ikx}\,\left\langle\left[J^{EM}_\mu(x),J^{EM}_\nu(0)\right]\right\rangle
= -G^<_{\nu\mu}(k)-G^>_{\nu\mu}(k)=\left(e^{\beta k^0}-1\right) G^<_{\nu\mu}(k)\quad.\label{GN}
\ee
Looking at the two definitions (\ref{GR}) and (\ref{GN}), one
has the following relation between $G^R$ and $G^N$,
\bear
G^R_{\mu\nu}(k)&=&-{1\over 2\pi}\int_{-\infty}^{+\infty}d\tilde k^0\,{1\over \tilde k^0-k^0-i\varepsilon} G^N_{\mu\nu}(\tilde k^0,\vec k)\nonumber\\&=&-{1\over 2\pi}{\cal P}\int_{-\infty}^{+\infty}d\tilde k^0\,{1\over \tilde k^0-k^0} G^N_{\mu\nu}(\tilde k^0,\vec k)
-{i\over 2}G^N_{\mu\nu}(k)\quad,\label{GRGN}
\eear
where ${\cal P}$ denotes Cauchy principal integral.
The relations (\ref{GN}) and (\ref{GRGN}) are the basic starting point of our discussion.

From translational invariance and hermiticity of $J^{EM}_\mu$, one easily derives
\be
G^N_{\mu\nu}(k)=-G^N_{\nu\mu}(-k) = G^{N *}_{\nu\mu}(k) \quad,
\ee
so that $G^N$ as well as $G^<$ and $G^>$ are hermitian matrices with $(\mu\nu)$ indices, which means importantly that
\be
\eta^{\mu\nu}G^N_{\mu\nu}(k)\quad,\quad \epsilon^\mu\epsilon^{\nu *}G^N_{\mu\nu}(k)\quad,
\ee
are real. Contracting (\ref{GRGN}) with $\eta^{\mu\nu}$ or $\epsilon^\mu\epsilon^{\nu *}$ therefore gives one that
\be
\eta^{\mu\nu}G^N_{\mu\nu}(k)=-2 \,{\rm Im}\left[\eta^{\mu\nu}G^R_{\mu\nu}(k)\right]\quad,\quad
\epsilon^\mu\epsilon^{\nu *}G^N_{\mu\nu}(k)=-2 \,{\rm Im}\left[\epsilon^\mu\epsilon^{\nu *}G^R_{\mu\nu}(k)\right]\quad.\label{NIM}
\ee
Note that
\be
{\rm Im}\left[\epsilon^\mu\epsilon^{\nu *}G^R_{\mu\nu}(k)\right]\neq \epsilon^\mu\epsilon^{\nu *}{\rm Im}\left[G^R_{\mu\nu}(k)\right]\quad,
\ee
in general, especially for circular polarization states where $\epsilon^\mu$ is complex. Note also that
\be
G^N_{\mu\nu}(k)\neq -2 \,{\rm Im}\left[G^R_{\mu\nu}(k)\right]\quad,\label{NR}
\ee
in general, as $G^N$ is hermitian but may not be real.
Only in the case of isotropic plasma, $G^N_{\mu\nu}(k)$ (and $G^{<,>}$) is diagonal and hence real, then the equality in (\ref{NR}) holds.

From (\ref{GRGN}) and (\ref{NIM}), one finally has
\bear
d\Gamma_\gamma &=& {d^3 k\over(2\pi)^3}{e^2\over 2 |\vec k|}{-2\over e^{\beta k^0}-1}{\rm Im}\,\left[\eta^{\mu\nu}G_{\mu\nu}^R(k)\right]\Bigg|_{k^0=|\vec k|}\quad,\label{basicformula}
\eear
and
\bear
d\Gamma_\gamma\left(\epsilon^\mu\right) &=& {d^3 k\over(2\pi)^3}{e^2\over 2 |\vec k|}{-2\over e^{\beta k^0}-1}{\rm Im}\,\left[\epsilon^\mu \epsilon^{\nu *}G_{\nu\mu}^R(k)\right]\Bigg|_{k^0=|\vec k|}\quad,\label{basicformula2}
\eear
which we use for our numerical computation of photon emission rates.

\section{Computational set-up in holographic QCD}

For our non-perturbative computation of photon emission rates in the presence of magnetic field at strong coupling,
we work in the Sakai-Sugimoto model; a holographic model which is closest to the large $N_c$ QCD with massless chiral quarks
in quenched approximation. The chiralities of massless quarks and the aspects of triangle anomaly manifest themselves in the model through
5-dimensional Chern-Simons terms in the action, which will play important roles in our results later.
The model was constructed originally via embedding $D8$/$\overline{D8}$ branes inside the holographic background generated by $N_c$ number of $D4$ branes. Roughly speaking, $N_c$ $D4$ branes describe Yang-Mills gluonic dynamics of large $N_c$ QCD, whereas $D8$/$\overline{D8}$ branes describe massless chiral quarks of left-handed/right-handed chiralities respectively.
Since electromagnetism couples to the vector-like global symmetry of these chiral quarks, we are interested only in the dynamics
of the 8-branes in quenched approximation for our photon emission rate computation.
The original model sits in a 10-dimensional geometry which is a warped product of 5-dimensional holographic space and
an extra internal space of $S^4\times S^1$, but for our purposes in this paper the internal space doesn't play
a meaningful role, and we will integrate out our action over it to have a reduction to the 5-dimensional holographic space from the start. 

The resulting $D8/\overline{D8}$ brane action is 
\bear
S_{D8/\overline{D8}}&=& -C R^{9\over 4}\int d^4x dU\,U^{1\over 4}\sqrt{-{\rm det}\left(g^*_{5D}+2\pi l_s^2 F\right)}\nonumber\\
&\mp& {N_c\over 96\pi^2}\int d^4x dU\, \epsilon^{MNPQR}A_M F_{NP}F_{QR}\quad,
\eear
where $\epsilon$-tensor is numerical and
\be
C={N_c^{1\over 2}\over 96 \pi^{11\over 2} g_s^{1\over 2} l_s^{15\over 2}}\quad,\quad R^3=\pi g_s N_c l_s^3\quad.
\ee
The sign of the Chern-Simons term in the second line encodes chirality that the 8-branes describe; the $D8$ brane with the upper sign describes the left-handed $U(N_F)_L$ chiral dynamics, and vice versa for $\overline{D8}$ and the right-handed chiral symmetry $U(N_F)_R$. In the deconfined phase of the theory that we are interested in here, 
the 5-dimensional geometry in Eddington-Finkelstein coordinate is given as  
\be
ds_{5D}^2=\left(U\over R\right)^{3\over 2}\left(-f(U) dt^2 +\sum_{i=1}^3 \left(dx^i\right)^2 \right) +2 dUdt\quad,\quad
f(U)=1-\left(U_T\over U\right)^3\quad,
\ee
which has a black-hole horizon at $U=U_T$ with
\be
U_T=R^3\left(4\pi T\over 3\right)^2\quad,
\ee
in terms of temperature $T$.
Since the final gauge theory observables are independent of $l_s$, one can conveniently choose $2\pi l_s^2\equiv 1$
which will be assumed throughout this paper. Sakai and Sugimoto determined the values of parameters in the model
by fitting to the pion decay constant and the $\rho$ meson mass with $N_c=3$. In units of GeV, this gives us
\be
C=0.0211\quad,\quad R^3=1.44\quad,
\ee
which defines the model without further free parameters.

In this background, 
the $D8$ and $\overline{D8}$ branes
do not meet each other, and their dynamics are independent of each other in leading $N_c$ approximation.  
The relation between these chiral fields on the 8-branes and the more conventional vector/axial fields is given by
\be
A_{V,A}={1\over 2}\left(\pm A_L+A_R\right)\quad,
\ee
where $A_{L,R}$ mean the gauge fields on $D8$ and $\overline{D8}$ branes respectively. Note that in terms of currents, this is equivalent to
\be
J_{V,A}=\pm J_L+J_R\quad.
\ee
Electromagnetism couples to the vector symmetry with the coupling strength $e$, which means practically for us
that we replace
\be
A_V\to e A_{EM}\quad,
\ee
where $A_{EM}$ is the electromagnetic gauge potential. In our situation of having constant magnetic field along say $x^3$
direction, this implies that we have
\be
F_{12}^V=eB\quad,\quad A_A\equiv 0\quad,
\ee
which is equivalent to having the same background magnetic fields on both $D8$ and $\overline{D8}$ branes,
\be
F_{12}^L=F_{12}^R =eB\quad.
\ee
It is easy to check that these constant background magnetic fields satisfy the equations of motion of 8-branes. 
 
Working in the Eddington-Finkelstein coordinate which is useful for computing retarded Green's functions
\be
ds_{5D}^2=\left(U\over R\right)^{3\over 2}\left(-f(U) dt^2 +\sum_{i=1}^3 \left(dx^i\right)^2 \right) +2 dUdt\quad,\quad
f(U)=1-\left(U_T\over U\right)^3\quad,
\ee
and using the identity
\be
\sqrt{{\rm det}(1+M)}=1+{1\over 2}{\rm tr}M +{1\over 8}\left({\rm tr} M\right)^2 -{1\over 4}{\rm tr} \left(M^2\right) +{\cal O}\left(M^3\right)\quad,
\ee
the quadratic expansion of the action in the presence of the background magnetic field $F_{12}^{(0)}=eB$ becomes
\bear
{\cal L}_{(2)}&=&{1\over 2}\Bigg(A(U)\left(F_{tU}\right)^2 -B(U)\left(F_{3U}\right)^2-
C(U)\left(\left(F_{1U}\right)^2+\left(F_{2U}\right)^2\right) -D(U)\left(\left(F_{13}\right)^2+\left(F_{23}\right)^2\right)\nonumber\\
&-&E(U)\left(F_{12}\right)^2 +2F(U)F_{t3}F_{3U}+2G(U)\left(F_{t1}F_{1U}+F_{t2}F_{2U}\right)\Bigg)\nonumber\\
&\mp&{N_c eB\over 8\pi^2}\left(A_t F_{3U}-A_3 F_{tU}+A_U F_{t3}\right)\quad,
\eear
where the coefficient functions are given by
\bear
A(U)&=&C U \sqrt{U^3+(eB)^2 R^3}\quad,\quad B(U)=C U f(U) \sqrt{U^3+(eB)^2 R^3} \quad,\nonumber\\
C(U)&=&C {U^4 f(U)\over \sqrt{U^3+(eB)^2 R^3}}\quad,\quad D(U)=C\left(R\over U\right)^3 {U^4\over\sqrt{U^3+(eB)^2 R^3}}\quad,
\nonumber\\
E(U)&=& C\left(R\over U\right)^3 {U^7\over\left(U^3+(eB)^2 R^3\right)^{3\over 2}}\quad,\quad
F(U)=C U \left(R\over U\right)^{3\over 2}\sqrt{U^3+(eB)^2 R^3}\quad,\nonumber\\
G(U)&=& C \left(R\over U\right)^{3\over 2}{U^4\over\sqrt{U^3+(eB)^2 R^3}}\quad.\label{coef}
\eear
The equations of motion derived from the above are
\bear
A(U)\partial_t F_{tU}-B(U)\partial_3 F_{3U}-C(U)\left(\partial_1 F_{1U}+\partial_2 F_{2U}\right)
+F(U)\partial_3 F_{t3}&&\nonumber\\
+G(U)\left(\partial_1 F_{t1}+\partial_2 F_{t2}\right)\pm{N_c eB\over 4\pi^2}F_{t3}&=&0\quad,\nonumber\\
\partial_U\left(A(U)F_{tU}\right)+F(U)\partial_3 F_{3U} +G(U)\left(\partial_1 F_{1U}+\partial_2 F_{2U}\right)\mp{N_c eB\over 4\pi^2}F_{3U}&=&0\quad,\nonumber\\
\partial_U\left(B(U)F_{3U}\right)-D(U)\left(\partial_1 F_{13}+\partial_2 F_{23}\right)+F(U)\partial_t F_{3U}-\partial_U\left(
F(U)F_{t3}\right)\mp{N_c eB\over 4\pi^2}F_{tU}&=&0\quad,\nonumber\\
\partial_U\left(C(U)F_{1U}\right)+D(U)\partial_3 F_{13} +E(U)\partial_2 F_{12}+G(U)\partial_t F_{1U}-\partial_U\left(G(U)F_{t1}\right)&=& 0\quad,\nonumber\\
\partial_U\left(C(U)F_{2U}\right)+D(U)\partial_3 F_{23}-E(U)\partial_1 F_{12} +G(U)\partial_t F_{2U}-\partial_U\left(G(U)F_{t2}\right)&=& 0\quad.\nonumber\\\label{eom}
\eear
Note that triangle anomaly represented by the 5D Chern-Simons term contributes to the equations of motion.
Physically, this is due to an interplay between vector and axial symmetries via chiral magnetic waves in the presence of the magnetic field. This
implies that the resulting photon emission rate encodes non-trivial effects from the triangle anomaly.

We work in the gauge $A_U=0$, and solve the equations of motion (\ref{eom}) for $A_\mu(U)$ in the frequency-momentum space
assuming the factor $e^{ikx}=e^{-i k^0 t+i\vec k\cdot \vec x}$.
Because there is a residual $SO(2)$ rotation symmetry on the transverse coordinates $(x^1,x^2)$, we can put $k^2=0$ 
without loss of generality. As we are interested in the elliptic flow, we write
\be
k^1=k^0 \sin\theta\quad,\quad k^3=k^0\cos\theta\quad,
\ee
satisfying the on-shell condition $k^0=|\vec k|\equiv \omega$ and showing the azimuthal angle $\theta$ from the direction of the magnetic field explicitly. The equations of motion then become
\bear
A(U)\partial_U A_t +\cos\theta B(U)\partial_U A_3+\sin\theta C(U)\partial_U A_1 -i\omega\cos\theta F(U)\left(A_3 +\cos\theta A_t\right)&&\nonumber\\
-i\omega\sin\theta G(U)\left(A_1 +\sin\theta A_t\right)\mp{N_c eB\over 4\pi^2}\left(A_3 +\cos\theta A_t\right)&=&0\quad,\nonumber\\
\partial_U\left(A(U)\partial_U A_t\right)+i\omega\cos\theta F(U)\partial_U A_3 +i\omega\sin\theta G(U)\partial_U A_1
\mp {N_c eB\over 4\pi^2}\partial_U A_3&=&0\quad,\nonumber\\
\partial_U\left(B(U)\partial_U A_3\right)-\omega^2\sin\theta D(U)\left(\sin\theta A_3-\cos\theta A_1\right)-i\omega F(U)\partial_U A_3&&\nonumber\\
-i\omega\partial_U\left(F(U)\left(A_3+\cos\theta A_t\right)\right)\mp {N_c eB\over 4\pi^2}\partial_U A_t &=& 0\quad,\nonumber\\
\partial_U\left(C(U)\partial_U A_1\right)+\omega^2\cos\theta D(U)\left(\sin\theta A_3-\cos\theta A_1\right)&&\nonumber\\
-i\omega G(U)\partial_U A_1 -i\omega \partial_U\left(G(U)\left(A_1+\sin\theta A_t\right)\right)&=&0\quad,\nonumber\\
\partial_U\left(C(U)\partial_U A_2\right)-\omega^2\cos^2\theta D(U) A_2 -\omega^2\sin^2\theta E(U) A_2 &&\nonumber\\
-i\omega G(U)\partial_U A_2 -i\omega \partial_U\left(G(U) A_2\right)&=& 0\quad.\nonumber\\\label{eomk}
\eear

As we are interested in the retarded Green's function of the global symmetries that the above 5-dimensional gauge fields describe, we need to establish a well-defined procedure of extracting information on the corresponding currents in the field theory side from the 5-dimensional profile of the solutions of the above equations. We follow standard steps of gauge/gravity
duality dictionary.
We first discuss how to extract expectation values of the symmetry currents in the field theory side from the 5-dimensional solutions.
To construct the boundary current expectation value $\langle J_\mu\rangle$ from the solution $A_\mu(U)$ of the above equations,
one has to study the near boundary asymptotics at $U\to\infty$ and perform a careful holographic renormalization.
The necessity of a careful holographic renormalization comes from the fact that we will turn on space-time varying external sources to find retarded Green's functions at finite frequency-momentum. In this situation, it 
generally happens that the renormalized boundary currents $\langle J_\mu\rangle$ are given not only by the coefficients of the
subleading terms in the asymptotic expansion, but also get some additional contributions from the external sources.
Only after combining these two contributions, the resulting currents $\langle J_\mu\rangle$ satisfy 
the correct conservation Ward identity. 

The near $U\to\infty$ asymptotics of $A_\mu$ are found to be
\be
A_\mu=A^{(0)}_\mu +{A^{(1)}_\mu\over U^{1\over 2}}+{A^{(2)}_\mu\over U}
+{A^{(3)}_\mu \log(U)\over U^{3\over 2}}+{\tilde A_\mu\over U^{3\over 2}}+\cdots\quad,\label{nearexp}
\ee
where we assume the dependence $e^{ikx}=e^{-i k^0 t+i\vec k\cdot \vec x}$ implicitly. We will consider a general $k^\mu$
for a while in our discussion of holographic renormalization.
$A^{(0)}_\mu$ is the external source and the subleading terms $A^{(1)}_\mu$, $A^{(2)}_\mu$, and $A^{(3)}_\mu$
are completely fixed by the external source $A^{(0)}_\mu$ in the following way,
\bear
A_t^{(1)}=0\quad,\quad A^{(3)}_\mu&=&0\quad,\nonumber\\
A^{(1)}_i=-2i R^{3\over 2}\left(k^0 A^{(0)}_i +k_i A^{(0)}_t\right)&=&2R^{3\over 2}F^{(0)}_{ti}\quad,\quad i=1,2,3\quad,\nonumber\\
A^{(2)}_t=-2 R^3\left(k^0 k^j A^{(0)}_j+k^jk_j A^{(0)}_t\right)&=&-2 R^3\partial_j F^{(0)}_{tj}\quad,\quad\nonumber\\
A^{(2)}_i=-2 R^3\left(k^j k_j A^{(0)}_i-k_i k^j A^{(0)}_j\right)&=&-2R^3 \partial_j F^{(0)}_{ij}\quad.\label{oriexp}
\eear
The fact that the coefficient of the logarithmic term, $A^{(3)}_\mu$, vanishes is 
special in this model, which would not be the case for a 5D gauge theory in asymptotic $AdS_5$ geometry.
The absence of logarithmic term translates to the absence of contribution to the conformal anomaly from the corresponding
global symmetry in the 4D gauge theory side. This can be understood because the 4D theory that this holographic model
supposedly describes is conformally non-invariant even in ultra-violet regime.
The piece $\tilde A_\mu$ is not determined by the external source, and it encodes a dynamical freedom of the expectation value
$\langle J_\mu\rangle$. It should be fixed by appropriate infrared boundary conditions, which in our case will be
the incoming boundary condition on the black hole horizon. In Eddington-Finkelstein coordinate we are working,
this simply requires regularity of $A_\mu(U)$ at the horizon $U=U_T$. In conjunction with the ultraviolet boundary
condition $A^{(0)}_\mu$, these two boundary conditions determine the solution $A_\mu(U)$ uniquely.

Although 
$\langle J_\mu\rangle$ should contain $\tilde A_\mu$ representing dynamical degrees of freedom, the full expression
for $\langle J_\mu\rangle$ involves additional contributions from (derivatives) of the external source $A^{(0)}_\mu$ as we mentioned above. To find these additional contributions, one normally goes through steps of holographic renormalization by
regularizing and subtracting divergences as we describe in the following.
For this purpose, it is more convenient to go to the diagonal metric frame,
\be
ds_{5D}^2=\left(U\over R\right)^{3\over 2}\left(-f(U) dt_*^2 +\sum_{i=1}^3 \left(dx^i\right)^2 \right)+\left(R\over U\right)^{3\over 2}{1\over f(U)} dU^2\quad,
\ee
by the coordinate transformation,
\be
t_*=t-\int_\infty^UdU'\,{1\over f(U')}\left(R\over U'\right)^{3\over 2}=t+{2 R^{3\over 2}\over U^{1\over 2}}+\cdots\quad.\label{trans}
\ee
Near $U\to\infty$ boundary, the temperature/magnetic field do not matter and there is 4D Lorentz symmetry one can use in discussing
asymptotics and expectation values in this frame.
The near boundary behavior of $A^*_\mu$ in the $A^*_U=0$ gauge in this frame is found to be (we put $*$ to mean the gauge fields in this diagonal metric frame)
\be
A^*_\mu=A^{(0)}_\mu -2 R^3\partial_\nu F^{(0)\,\nu}_\mu{1\over U}+{\tilde A^*_\mu\over U^{3\over 2}}+\cdots\quad.\label{exp1}
\ee
Note the absence of $U^{-{1\over 2}}$ term as well as the logarithmic term. Upon inserting the above to the 5D holographic action\footnote{One can neglect effects from temperature and magnetic field in near $U\to\infty$ boundary as they are sufficiently subleading. The 5D Chern-Simons term is also irrelevant in divergences and minimal counterterms and we will  consider it only at the end in the final results.}
\be
S_{5D}=-{C\over 4}\int d^4x dU\,\left(2 U^{5\over 2} F_{\mu U}F^{\mu U}+{R^3\over U^{1\over 2}}F_{\mu\nu}F^{\mu\nu}\right)\quad,
\ee
there appears a divergence,
\be
S_{5D}^{reg}\sim -{C\over 2} R^3 U_\infty^{1\over 2}\int d^4 x\, F^{(0)}_{\mu\nu}F^{(0)\mu\nu} \quad,
\ee
where we regularize the divergence at $U=U_\infty$. The counterterm at $U=U_\infty$ is hence chosen to be
\be
S^{ct}=+{C\over 2} R^3 U_\infty^{1\over 2}\int d^4 x\, F_{\mu\nu}F^{\mu\nu}\Bigg|_{U=U_\infty} \quad,\label{ct}
\ee
to cancel the divergence minimally. Then the renormalized current expectation value from $S^{ren}=S^{reg}+S^{ct}$ is
\bear
\langle J_\mu \rangle &=& \lim_{U_\infty\to\infty}\left(-C U_\infty^{5\over 2}\partial_U A_\mu(U_\infty)+2C R^3
U_\infty^{1\over 2}\partial_\nu F^{\mu\nu}(U_\infty)\right)={3\over 2}C \tilde A^*_\mu\quad.\label{result}
\eear
The result is simple without further contributions from the source; this is special in this model, which is related to
the absence of logarithmic term in the expansion (\ref{exp1}).

The 5D Chern-Simons term does not introduce further divergences, so the minimal counterterm (\ref{ct}) remains unchanged.
It only modifies the regularized action $S^{reg}$ by (recall our $A^*_U=0$ gauge)
\be
\delta S^{reg}=\mp {N_c eB\over 12\pi^2}\int d^4x dU\,\epsilon^{12\mu\nu} A_\mu F_{\nu U}\mp{N_c\over 24\pi^2}
\int d^4x dU\,\epsilon^{\mu\nu\alpha\beta}A^B_\mu F_{\nu\alpha}F_{\beta U}
\quad,
\ee
with $\epsilon^{t123}=1$, and $A^B$ is the gauge field responsible for the background magnetic field, $A^B_{1,2}=\mp{eB\over 2} x^{2,1}$. The correction to $\langle J_\mu \rangle$ from this is thus
\be
\delta \langle J_\mu \rangle=\mp{N_c eB\over 12\pi^2}\epsilon_{12\mu\nu}A^{(0)\nu}\pm{N_c\over 24\pi^2}
\epsilon_{\mu\nu\alpha\beta}A^{B\nu}F^{(0)\alpha\beta}\quad.\label{add}
\ee
When we consider both $D8$ and $\overline{D8}$ branes together in discussing vector/axial symmetries,
one can still add a further finite piece in the counterterm $S^{ct}$, called the Bardeen counterterm, to ensure
the conservation of vector symmetry, and this will modify $\langle J_\mu \rangle$ additionally \cite{Rebhan:2009vc}.

Although one may deal with them carefully to get the final results correctly, 
for our problem of computing vector-vector Green's functions it turns out that these modifications from (\ref{add}) and the Bardeen counterterm simply drop in the final results.
Recall that the vector current is a simple sum of left- and right-handed currents, so that the modification (\ref{add})
gives rise to
\be
\delta \langle J_V\rangle= {N_c eB\over 6\pi^2}\epsilon_{12\mu\nu}A_A^{(0)\nu}-{N_c\over 12\pi^2}\epsilon_{\mu\nu\alpha\beta}
A^{B\nu}F^{(0)\alpha\beta}_A\quad,\label{mod}
\ee
where $A_A^{(0)}\equiv {1\over 2}\left(-A_L^{(0)}+A_R^{(0)}\right)$ is the source for the axial current $J_A=-J_L+J_R$.
The modification from the Bardeen counterterm is also easily found to be
\be
\delta \langle J_V\rangle= {N_c eB\over 3\pi^2}\epsilon_{12\mu\nu}A_A^{(0)\nu}+{N_c\over 12\pi^2}\epsilon_{\mu\nu\alpha\beta}
A^{B\nu}F^{(0)\alpha\beta}_A\quad.\label{bardeen}
\ee
Both (\ref{mod}) and (\ref{bardeen}) give the response of $\langle J_V\rangle$ proportional to the external axial source $A_A$ only, so that they don't contribute to the vector-vector Green's functions.

\begin{figure}[t]
	\centering
	\includegraphics[width=12cm]{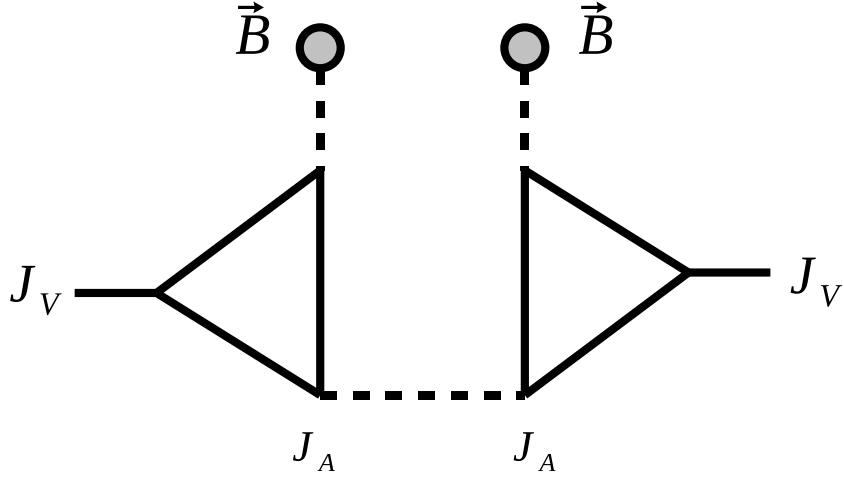}
		\caption{Contribution of triangle anomaly to the vector-vector Green's functions in the presence of magnetic field.\label{fig0}}
\end{figure}

Based on this observation, we will use the formula (\ref{result}) for the current expectation values for simplicity, without
worrying about further corrections from the 5D Chern-Simons term.
We stress that the presence of the 5D Chern-Simons term does affect our results 
via bulk equations of motion (\ref{eom}); it affects the dynamical coefficient $\tilde A_\mu$ and hence the current expectation
values through equations of motion. 
This can easily be seen by adding and subtracting (\ref{eom}) for upper/lower signs of Chern-Simons term (corresponding to $D8$/$\overline{D8}$ branes for left/right-handed chiralities) to get
\bear
A(U)\partial_t F_{tU}^{V,A}-B(U)\partial_3 F_{3U}^{V,A}-C(U)\left(\partial_1 F_{1U}^{V,A}+\partial_2 F_{2U}^{V,A}\right)
+F(U)\partial_3 F_{t3}^{V,A}&&\nonumber\\
+G(U)\left(\partial_1 F_{t1}^{V,A}+\partial_2 F_{t2}^{V,A}\right)-{N_c eB\over 4\pi^2}F_{t3}^{A,V}&=&0\quad,\nonumber\\
\partial_U\left(A(U)F_{tU}^{V,A}\right)+F(U)\partial_3 F_{3U}^{V,A} +G(U)\left(\partial_1 F_{1U}^{V,A}+\partial_2 F_{2U}^{V,A}\right)+{N_c eB\over 4\pi^2}F_{3U}^{A,V}&=&0\quad,\nonumber\\
\partial_U\left(B(U)F_{3U}^{V,A}\right)-D(U)\left(\partial_1 F_{13}^{V,A}+\partial_2 F_{23}^{V,A}\right)+F(U)\partial_t F_{3U}^{V,A}&&\nonumber\\-\partial_U\left(
F(U)F_{t3}^{V,A}\right)+{N_c eB\over 4\pi^2}F_{tU}^{A,V}&=&0\quad,\nonumber\\
\partial_U\left(C(U)F_{1U}^{V,A}\right)+D(U)\partial_3 F_{13}^{V,A} +E(U)\partial_2 F_{12}^{V,A}+G(U)\partial_t F_{1U}^{V,A}-\partial_U\left(G(U)F_{t1}^{V,A}\right)&=& 0\quad,\nonumber\\
\partial_U\left(C(U)F_{2U}^{V,A}\right)+D(U)\partial_3 F_{23}^{V,A}-E(U)\partial_1 F_{12}^{V,A} +G(U)\partial_t F_{2U}^{V,A}-\partial_U\left(G(U)F_{t2}^{V,A}\right)&=& 0\quad,\nonumber\\\label{eomAV}
\eear
where $A^{V,A}={1\over 2}\left( \pm A_L+A_R\right)$. It is clear that axial components are necesssarily excited in computing
vector-vector Green's functions due to Chern-Simons term. A diagramatic representation is given in Figure \ref{fig0}. 
Therefore, the effects from triangle anomaly to our photon emission observables
are totally dynamical, and are not sensitive to the issues of additional boundary contributions.
In retrospect this makes good sense because at the end we are dealing with only vector-like observables of electromagnetism which should be unambiguously defined. 

What remains to find $\langle J_\mu\rangle$ in the original Eddington-Finkelstein coordinate is to 
perform the coordinate transformation of (\ref{trans}) on $A_\mu (U)$, so that one can express $\tilde A^*_\mu$
in terms of $\tilde A_\mu$ and $A^{(0)}_\mu$.
One has to be careful about our gauge choices $A_U=0$ and $A^*_U=0$; starting from $A_\mu(U)$ and performing
the coordinate transformation (\ref{trans}), the resulting $A^*_U$ is not zero and one needs to do a further
gauge transformation to remove $A_U^*$. Note that $\partial_t=\partial_{t^*}$, but $\partial_U\neq \partial_{U^*}$.
The computation is straightforward and the result is
\bear
\tilde A^*_t&=&\tilde A_t+{8\over 3} R^{9\over 2}\partial_t\partial_jF_{tj}^{(0)}\quad,\nonumber\\
\tilde A^*_i&=& \tilde A_i +4R^{9\over 2}\left(\partial_t\partial_j F^{(0)}_{ij}+{2\over 3}\partial_t^2F_{ti}^{(0)}
-{1\over 3} \partial_i\partial_j F^{(0)}_{tj}\right)\quad,\label{result2}
\eear
from which one obtains the current expectation values via $\langle J_\mu\rangle={3\over2}C \tilde A^*_\mu$ as in (\ref{result}).
One can check that the first equation in (\ref{eom}) gives one the correct chiral Ward identity (up to the additional
boundary contributions and the Bardeen counterterm discussed above)
\be
\partial_\mu \langle J^\mu\rangle =\pm {N_c eB\over 4\pi^2}F_{t3}^{(0)}\quad,\label{ward}
\ee
in the presence of external chiral gauge fields $eB$ and $A^{(0)}$.

Once we know how to extract the expectation values from the solutions as above, we can easily compute the retarded Green's
functions in the following way. The leading component $A^{(0)}_\mu$ in the near boundary expansion (\ref{nearexp})
is interpreted in the QCD field theory side as an external gauge potential coupling to the (chiral) current $J_{L,R}^\mu$.
The chirality depends on the sign of the Chern-Simons term, or equivalently on whether we are looking at $D8$ or $\overline{D8}$ brane. 
By demanding incoming boundary condition at the horizon, which is simply a regularity at the horizon in our Eddington-Finkelstein coordinate, the solution is uniquely determined by this external source $A^{(0)}_\mu$;
it is clear that the solution linearly depends on $A^{(0)}_\mu$.
One then obtains the expectation value of chiral current from the solution by using (\ref{result}) and (\ref{result2}).
As the result is linear in $A^{(0)}_\mu$, one writes
\be
\langle J_\mu \rangle=-G^{R\,\,\,\,\nu}_{\mu}(k)A^{(0)}_\nu\quad,\label{green}
\ee
which gives one the retarded Green's function $G^R_{\mu\nu}$ by the definition of Kubo's formulation of real-time response functions. 
For the upper sign of Chern-Simons term in (\ref{eom}) ( that is from $D8$ brane), one obtains 
the Green's function of left-handed chiral current $G^R_{LL}$, and vice versa for the lower sign (that is $\overline{D8}$ brane)
and the right-handed $G^R_{RR}$. The desired vector-vector Green's function for the photon emission rate is then
\be
G^R_{VV}=G^R_{LL}+G^R_{RR}\quad.
\ee
Because the two chiral Green's functions are simply related to each other by
\be
G_{LL}(eB)=G_{RR}(-eB)\quad,
\ee
or equivalently
\be
G_{LL}(\theta)=G_{RR}(\pi-\theta)\quad,
\ee
in terms of azimuthal angle $\theta$ with respect to the magnetic field,
it is enough to compute $G_{LL}$ only.

We numerically solve the equations of motion (\ref{eom}), or more precisely (\ref{eomk}), 
to compute the retarded Green's functions from the relation (\ref{green}).
Instead of solving the equations with a given $A^{(0)}_\mu$ from the ultraviolet (UV) boundary $U\to\infty$,
it is numerically much easier to solve them from the horizon $U=U_T$ toward the UV boundary, starting with a regularity boundary condition at the horizon.
Inspecting the equations of motion (\ref{eomk}) near the horizon $U=U_T$, the regularity condition
uniquely fixes the derivatives $\partial_U A_\mu(U_T)$ and $\partial^2_U A_\mu(U_T)$ in terms of the value
at the horizon $A_\mu(U_T)$, which allows one to start solving the equations (\ref{eomk}) from the horizon.
Practically, we use $\partial_U A_\mu(U_T)$ and $\partial^2_U A_\mu(U_T)$ to proceed a small step
to $U=U_T+\epsilon$ with a small number $\epsilon=0.01$, and start our numerical solving the equations (\ref{eomk}) from that point until a UV cutoff $U=U_{max}$.
Given this solution, one obtains $A^{(0)}_\mu$ and $\tilde A_\mu$ by comparing with the near boundary expansion (\ref{nearexp}) and (\ref{oriexp}) at the position $U=U_{max}$. More precisely, we solve the following linear system of equations for $(A^{(0)},\tilde A)$,
\bear
A_\mu(U_{max})&=&A_\mu^{(0)}+{A_\mu^{(1)}\over U_{max}^{1\over 2}}+{A_\mu^{(2)}\over U_{max}}+{\tilde A_\mu \over U_{max}^{3\over 2}}+{A_\mu^{(4)}\over U_{max}^{2}}\quad,\nonumber\\
\left(\partial_U A_\mu\right)(U_{max})&=& -{1\over 2}{A_\mu^{(1)}\over U_{max}^{3\over 2}}-{A_\mu^{(2)}\over U_{max}^{2}}-{3\over 2}{\tilde A_\mu\over U_{max}^{5\over 2}}-2{A_\mu^{(4)}\over U_{max}^{3}}\quad,\label{numericalexp}
\eear
where the left-hand sides are given by the numerical solution, and $A^{(1,2,4)}_\mu$ are linearly given by $A^{(0)}_\mu$ as in (\ref{oriexp}).
Note that we have included one more term, $A^{(4)}$, in the above
expansion on the right-hand side than in (\ref{nearexp}) for a better numerical precision of extracting $(A^{(0)},\tilde A)$
from the above. The expressions for $A^{(4)}_\mu$  can be easily found and we skip their explicit expressions.

From $(A^{(0)},\tilde A)$, one then constructs $(A^{(0)},\langle J\rangle)$ using (\ref{result}) and (\ref{result2}).
Since it is clear that the results are linearly dependent on the horizon data $A_\mu(U_T)$ that we start with,
one writes
\be
A^{(0)}_\mu = {\cal S}_\mu^{\,\,\,\,\nu}A_\nu(U_T)\quad,\quad \langle J_\mu\rangle= {\cal R}_\mu^{\,\,\,\,\nu}A_\nu(U_T)\quad,
\ee
with two matrices ${\cal S}$ and ${\cal R}$. One easily computes these two matrices by performing the above described numerical
procedures for each four unit vectors of $A_\mu(U_T)$. Once they are found, one relates $(A^{(0)},\langle J\rangle)$ directly by
\be
\langle J_\mu\rangle = {\cal R}_\mu^{\,\,\,\,\nu}\left({\cal S}^{-1}\right)_\nu^{\,\,\,\,\,\,\alpha} A_\alpha^{(0)}
=\left({\cal R}\cdot{\cal S}^{-1}\right)_\mu^{\,\,\,\,\,\,\nu} A_\nu^{(0)}\equiv -G^{R\,\,\,\,\nu}_{\mu}A^{(0)}_\nu\quad,
\ee
so that the retarded Green's function is finally computed as $G^R=-{\cal R}\cdot{\cal S}^{-1}$ in a matrix form.
In the total photon emission rate formula (\ref{basicformula}), one needs the trace of the Green's function and
\be
\eta^{\mu\nu} G^R_{\mu\nu}= G_\mu^{R\,\,\,\,\mu}=-{\rm tr}\left({\cal R}\cdot{\cal S}^{-1}\right)\quad,
\ee
which is particularly simple in this matrix form. However, the full matrix Green's function contains much more information
than the simple trace, such as polarization states of emitted photons. We will discuss polarization asymmetry of the photons in  section \ref{pol},
which we propose to be an important new experimental signature of triangle anomaly in heavy-ion experiments.

From explicit expressions for $A^{(1,2,4)}_\mu$, one notes that the expansion parameter in the near boundary series (\ref{numericalexp}) is 
\be
x={R^{3\over 2}\omega\over U_{max}^{1\over 2}}\quad,
\ee
so that the larger the energy $\omega$ is, the bigger $U_{max}$ should be for a good numerical precision
of $A^{(0)}_\mu$ and $\tilde A_\mu$ (and hence of the retarded Green's functions) from solving the equations (\ref{numericalexp}).
From (\ref{numericalexp}), the error on $(A^{(0)},\tilde A)$ from neglecting higher order terms on the right-hand side is of order $x^2$.
In our numerical analysis, we have controlled $x^2$ to be much less than 0.01, so that our numerical results
should be reliable up to better than $1\%$. Also,
the structure of equations of motion (\ref{eom}) is such that the derivative of the first equation in (\ref{eomk}) 
is equivalent to the remaining four equations. Therefore, once the first equation is satisfied at one point in $U$,
say at $U=U_T$, it should remain satisfied for all $U$ if one solves the other equations correctly.
In other words, it is a consistent contraint equation of the system. It serves in fact useful in our numerical analysis
by providing a test of the numerical precision, and we have checked that our solutions satisfy it with much better precision than $1\%$. Another independent check is the chiral Ward identity (\ref{ward}) which also holds to a good precision.
This gives us the conservation Ward identities of our final vector-vector Green's function,
\be
k^\mu G^{R\,\,\,\,\nu}_\mu(k)=G^{R\,\,\,\,\nu}_\mu(k) k_\nu=0\quad,
\ee
which are checked to be true with a very good precision.


\section{Elliptic ($v_2$) and quadrupole ($v_4$) flows of photons}

\begin{figure}[t]
	\centering
	\includegraphics[width=5cm]{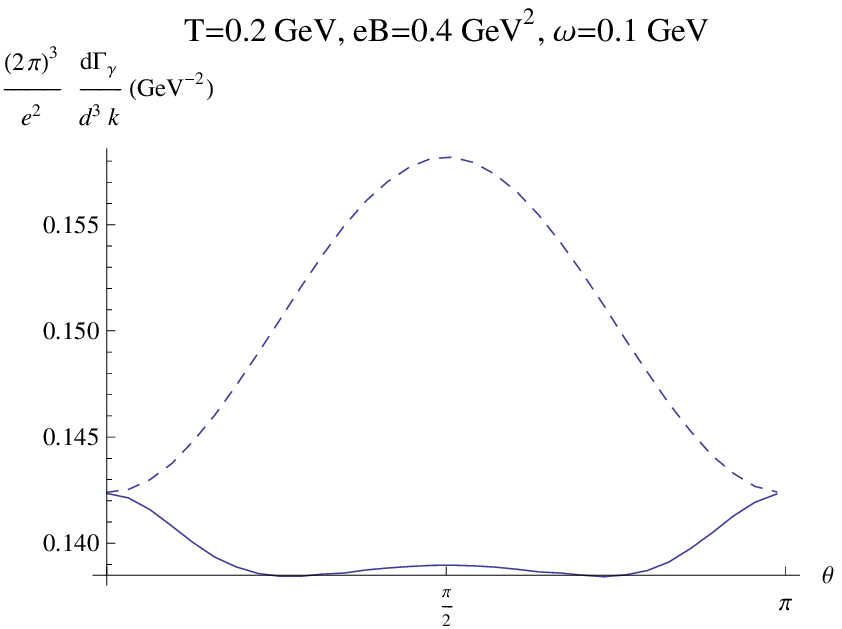}
\includegraphics[width=5cm]{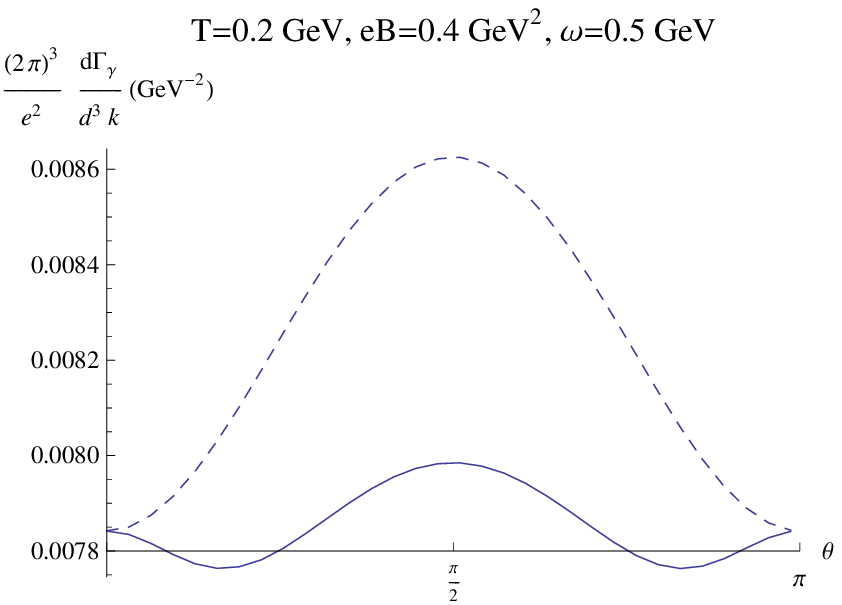}
\includegraphics[width=5cm]{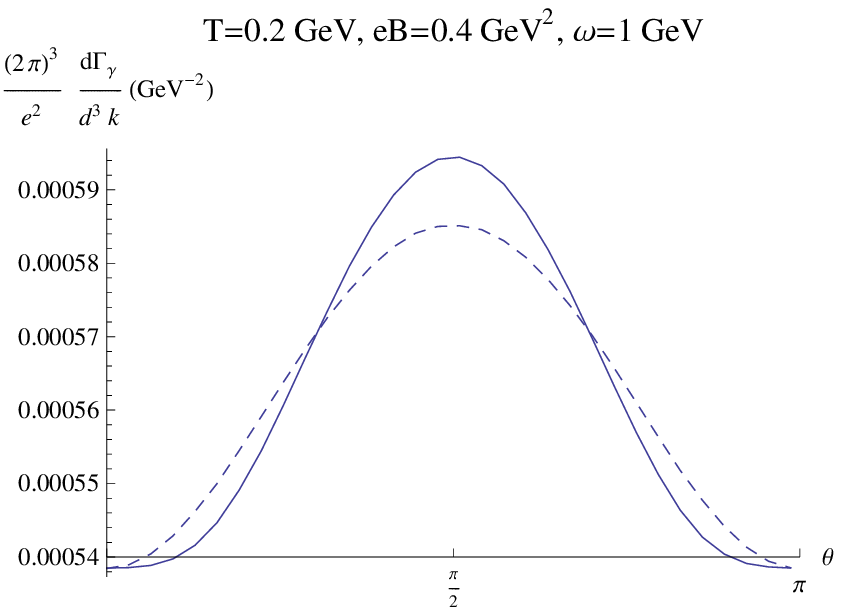}
		\caption{ Azimuthal angle dependence of photon emission rates for three different energies $\omega=0.1,0.5,1$ GeV with temperature $T=0.2$ GeV and magnetic field $eB=0.4$ ${\rm GeV}^2$. The dashed curves are the results when the Chern-Simons term is turned off.\label{fig1}}
\end{figure}
\begin{figure}[t]
	\centering
	\includegraphics[width=7cm]{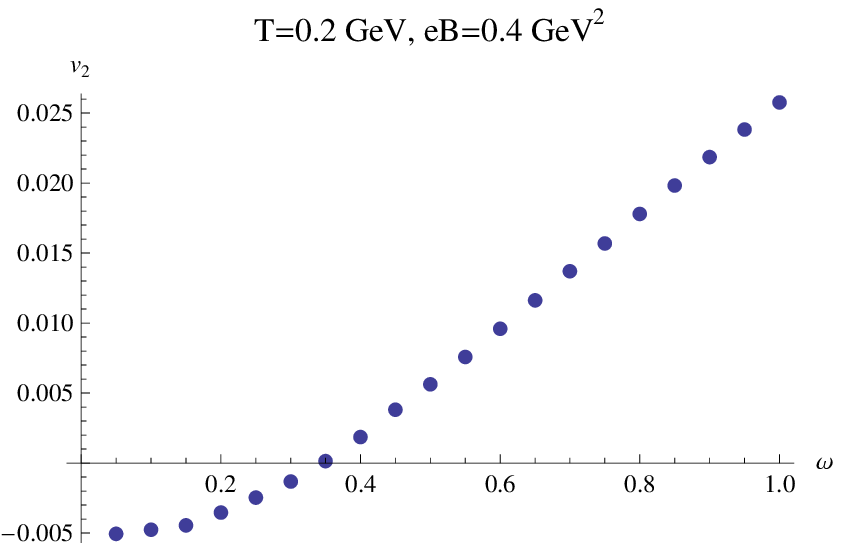}
\includegraphics[width=7cm]{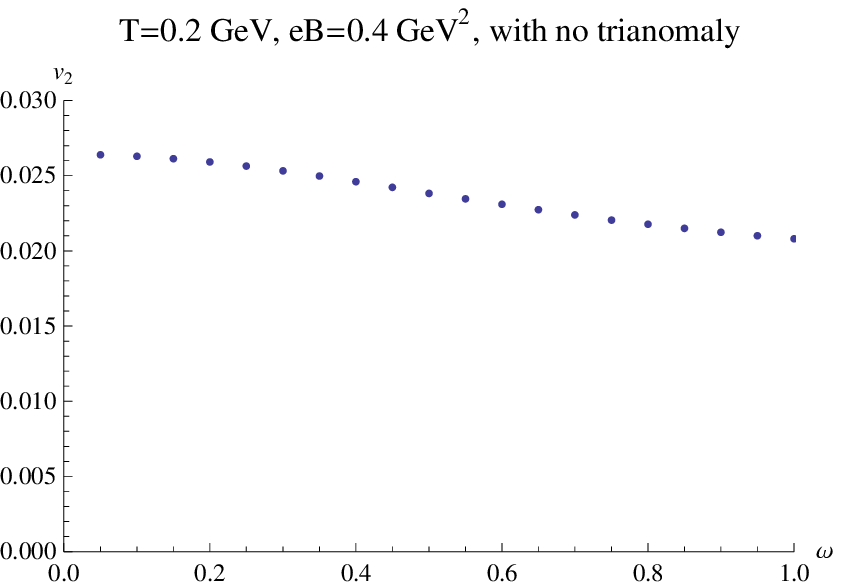}
		\caption{Elliptic flow $v_2$ versus photon energy $\omega$ for $T=0.2$ GeV and $eB=0.4$ ${\rm GeV}^2$. The right plot is the result without Chern-Simons term (triangle anomaly).\label{fig2}}
\end{figure}
\begin{figure}[t]
	\centering
	\includegraphics[width=7cm]{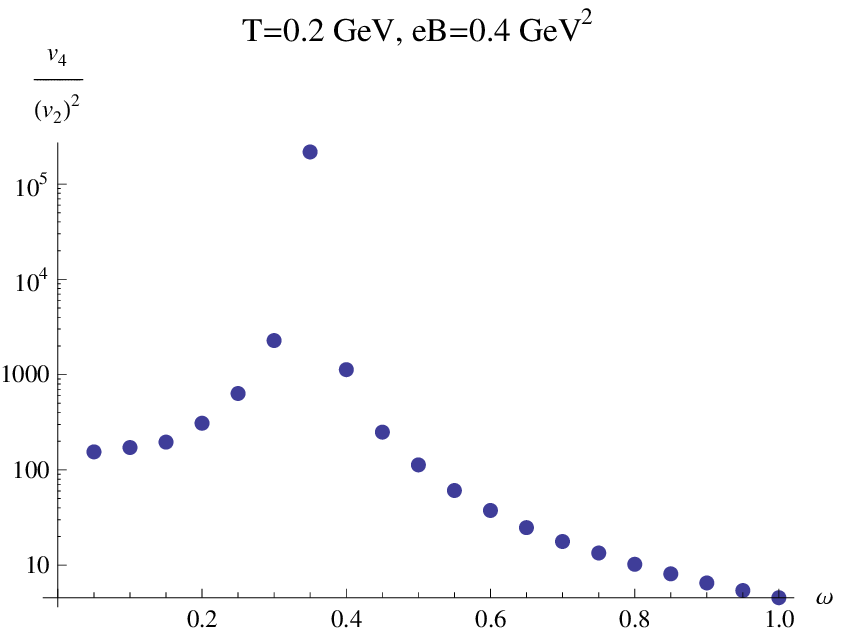}
\includegraphics[width=7cm]{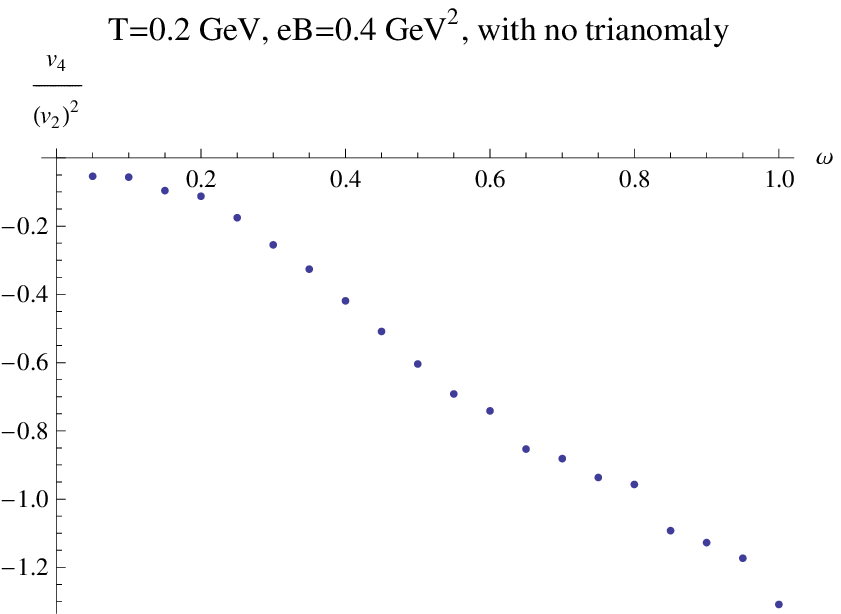}
		\caption{The ratio $v_4/(v_2)^2$ versus photon energy $\omega$ for $T=0.2$ GeV and $eB=0.4$ ${\rm GeV}^2$. The right plot is the result without Chern-Simons term (triangle anomaly).\label{fig4}}
\end{figure}
\begin{figure}[t]
	\centering
	\includegraphics[width=7cm]{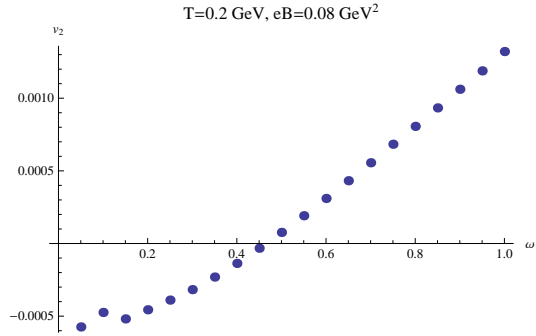}
		\caption{Elliptic flow $v_2$ versus photon energy $\omega$ for $T=0.2$ GeV and $eB=4m_\pi^2=0.08$ ${\rm GeV}^2$.\label{fig8}}
\end{figure}

With the computations described in the previous section, we present our results for the azimuthal dependence of the photon emission rates in the presence of the magnetic field.
We introduce the mode expansion in the azimuthal angle $\theta$ with respect to the magnetic field direction as
\be
{d\Gamma_\gamma\over d^3 k }(\theta)=\Gamma_0\left(1-2 v_2 \cos(2\theta)+2 v_4\cos(4\theta)+\cdots\right)\quad,
\ee
with the elliptic flow $v_2$ and the quadrupole flow $v_4$. 
The negative sign in front of $v_2$ in the above definition is due to the relation
\be
\phi={\pi\over 2}-\theta\quad,
\ee
between $\theta$ and the more conventional angle $\phi$ from the reaction plane (note that magnetic field is perpendicular to
the reaction plane).
In Figure \ref{fig1}, we show some exemplar plots of azimuthal dependence of photon emission rates ${d\Gamma_\gamma\over d^3 k }(\theta)$ for different energies.
We take $T=0.2$ GeV, $eB=0.4\,{\rm GeV}^2$ for an illustrative purpose (we will also consider more realistic values of the magnetic field later).
For useful comparison, we also include results obtained after dropping the Chern-Simons term (the dashed curves), which shows the importance of triangle (chiral) anomaly
in the results, especially for low energy $\omega\le 1$ GeV. As is clear from the plots, the angular dependence is drastically affected by the triangle anomaly for low energy regime (the plots in the left), and the mode expansion becomes more complicated than simply being characterized by an elliptic flow. Especially, the size of the quadrupole moment $v_4$ is comparable in this regime. We expect that these modifications of the azimuthal dependence are due to the existence of the chiral magnetic wave modes that affect the retarded Greens functions via its pole structure
\be
{1\over (\omega- v_\chi k \cos\theta )}+{1\over (\omega+ v_\chi k \cos\theta )}\,,
\ee
where $v_\chi$ is the velocity of the chiral magnetic wave and the two contributions are from left- and right-handed chiral magnetic waves respectively.
We plot $v_2$ (in Figure~\ref{fig2}) and the ratio $v_4/(v_2)^2$ (in Figure~\ref{fig4}) to highlight this effect, for example the violation of the usual scaling $v_4\sim v_2^2$. In the elliptic flow $v_2$, we notice that the elliptic flow can even be negative for the low energy $\omega\le 0.3$ GeV due to the effect from triangle anomaly.
Current experiments give data only for $\omega\ge 1$ GeV, so that the experimental relevance of this observation is not high at the moment, but it may become important in the future.
In Figure \ref{fig8}, we plot the elliptic flow $v_2$ for a realistic value of the magnetic field $eB=4m_\pi^2\approx 0.08\,{\rm GeV}^2$.
The general trends are the same as we describe in the above, although we observe that the overall magnitude of the elliptic flow is rather small, $v_2\sim 10^{-3}$.
We should take this as a result from the strong coupling computation via gauge/gravity correspondence.  The particular reason why the strong coupling
gives such a small azimuthal imbalance of the photon emission even with a large strength of the magnetic field is not completely clear to us, and might deserve a further study. 

\section{Polarization asymmetry of photons \label{pol}}

\begin{figure}[t]
	\centering
	\includegraphics[width=10cm]{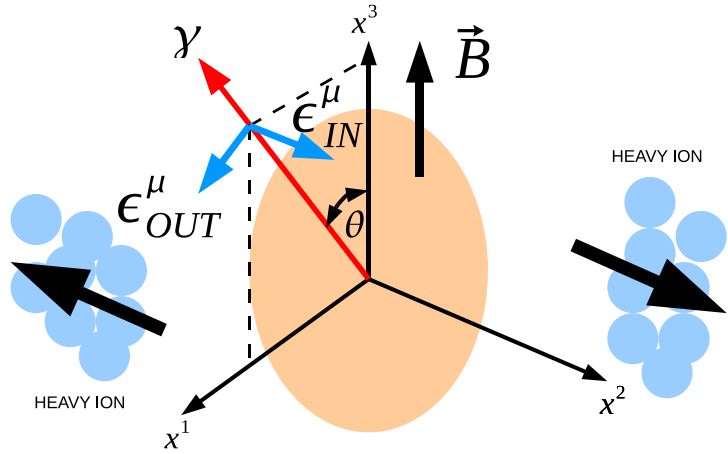}
		\caption{Definition of in- and out-plane polarizations.\label{fig6}}
\end{figure}

Our computational framework in gauge/gravity correspondence is able to describe the photon emissions with specific polarization states of the photons.
With our previous choice of photon momentum
\be
k^\mu=\omega\left(1,\sin\theta,0,\cos\theta\right)\quad,
\ee
we define the in-plane polarization with 
\be
\epsilon^\mu_{\rm IN}=\left(0,0,1,0\right)\quad,
\ee
and the out-plane polarization with
\be
\epsilon^\mu_{\rm OUT}=\left(0,\cos\theta,0,-\sin\theta\right)\quad.
\ee
See Figure \ref{fig6} for a schematic explanation of our definition of the polarization states. 
\begin{figure}[t]
	\centering
	\includegraphics[width=5cm]{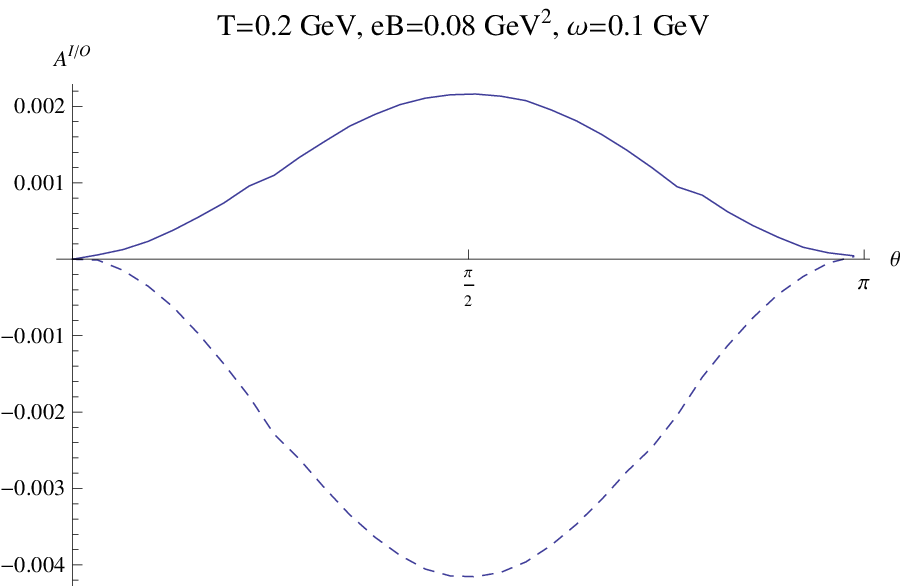}
\includegraphics[width=5cm]{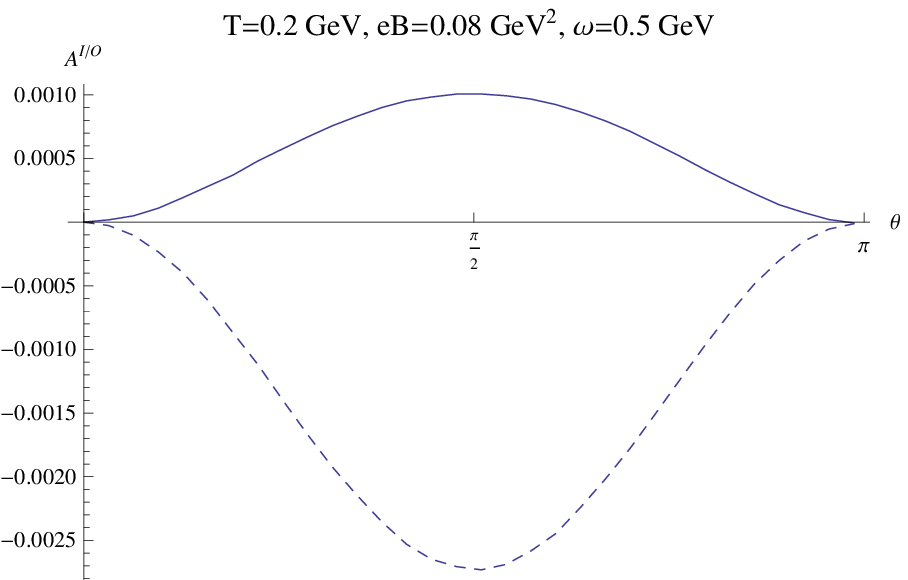}\includegraphics[width=5cm]{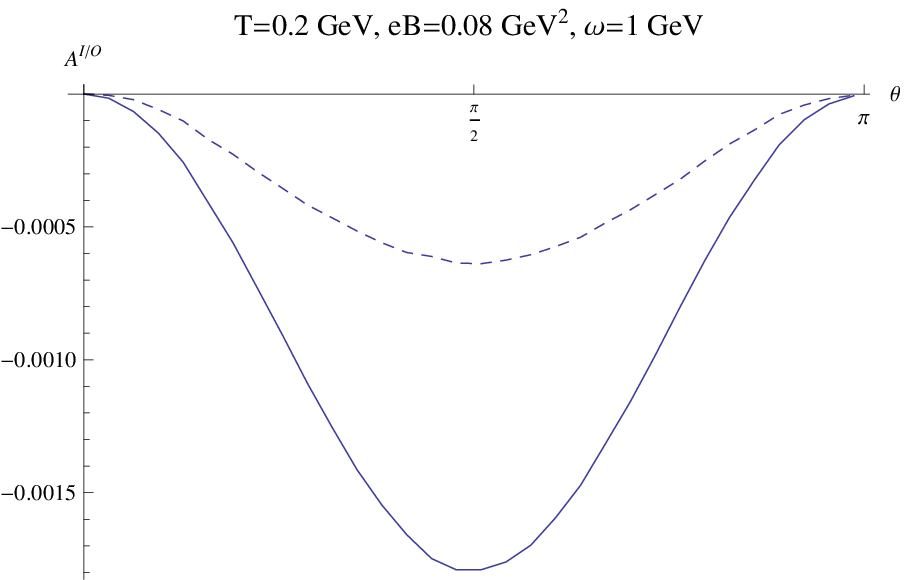}
		\caption{In/out-plane polarization asymmetry $A^{{\rm I/O}}$ for three different energies $\omega=0.1,0.5,1$ GeV with temperature $T=0.2$ GeV and magnetic field $eB=4m_\pi^2=0.08$ ${\rm GeV}^2$. The dashed curves are the results without Chern-Simons term (triangle anomaly). \label{fig7}}
\end{figure}
We then define a new observable, the ``in/out-plane polarization asymmetry'' $A^{\rm I/O}$ by 
\be
A^{\rm I/O}={{d\Gamma_\gamma\over d^3 k}(\epsilon_{\rm IN})-{d\Gamma_\gamma\over d^3 k}(\epsilon_{\rm OUT})\over
{d\Gamma_\gamma\over d^3 k}(\epsilon_{\rm IN})+{d\Gamma_\gamma\over d^3 k}(\epsilon_{\rm OUT})}\quad.
\ee
In Figure \ref{fig7}, we plot the azimuthal dependence of the in/out-plane polarization asymmetry $A^{\rm I/O}$ for different energies.  
As before, we observe that the Chern-Simons term affects the results in a fundamental way for low energy regime $\omega\le 1$ GeV, and we expect that the chiral magnetic wave pole
to be responsible for this modification. The general trend is that $A^{\rm I/O}$ is positive for low energy regime $\omega \le 1$ GeV, whereas it becomes negative for higher energies $\omega \ge 1$ GeV. Although current experiments seem to find it difficult to produce data for the photon polarization, our observation may become relevant in  more refined experimental measurements in the future.

\newpage

\vskip 1cm \centerline{\large \bf Acknowledgement} \vskip 0.5cm

We thank Dima Kharzeev for many discussions that led to an initial collaboration on this work.
We also thank Gokce Basar, Frasher Loshaj, Kiminad Mamo, Larry McLerran, Edward Shuryak, Vladimir Skokov, Todd Springer, Misha Stephanov, Derek Teaney, Kirill Tuchin, Yi Yin, and Ismail Zahed for discussions.

 \vfil

\end{document}